\documentstyle[preprint,revtex]{aps}
\begin{document}
\draft
\preprint{}
\begin{title}
A VARIATIONAL APPROACH TO BOUND STATES\\ IN QUANTUM FIELD THEORY
\end{title}
\author{H. Mishra and S.P. Misra}
\begin{instit}
Institute of Physics, Bhubaneswar-751005 India
\end{instit}
\vspace{3cm}
\begin{abstract}
We consider here in a toy model an approach to bound state problem in a
nonperturbative manner using equal time algebra for the
interacting field operators. Potential is replaced by
offshell bosonic quanta  inside the bound state of nonrelativistic particles.
The bosonic dressing is determined through energy minimisation, and mass
renormalisation is carried out in a nonperturbative manner. Since the
interaction is through a scalar field, it does not include
spin effects. The model however nicely incorporates an
intuitive picture of hadronic bound states in which the gluon
fields dress the quarks providing the binding between them and
also simulate the gluonic content of hadrons in deep inelastic
collisions.
\end{abstract}
\pacs{}

\section {\bf Introduction}

\label{sec:intro}
We have recently considered a nonperturbative variational method for structure
problems where the basic inputs are equal time algebra for
the interacting field operators and a variational ansatz as
may be appropriate for the specific dynamical situation. This
has been applied to vacuum problem in Gross Neveu model \cite{sah4} and
in quantum chromodynamics  for finite temperature and
baryon densities \cite{ahss}. It
has further been applied to chiral symmetry breaking \cite{ahss1} where a
new insight for vacuum structure with low energy hadronic properties
is gained \cite{amspm} as well as for the ground state of symmetric
nuclear matter \cite{nm} and neutron matter \cite{nn}. We shall extend the
method here to the consideration of bound states.

We have earlier considered this in a formal way \cite{spm3} and in nuclear
physics for deuteron \cite{deut}.
In the present paper we shall extend this method to include
renormalisation effects in a nonperturbative manner
while considering dressing of fermions, and then examine
quantitatively the boson content of the bound state as well as the momentum
carried by these offshell quanta.

We organise the paper as follows. In section {\bf 2} we
consider the Hamiltonian of two fermions interacting through Yukawa coupling
along with a counter term. We next obtain the energy of a single
dressed fermion and identify a nonperturbative  renormalisation
procedure for the present approximation scheme.
In section {\bf 3} we construct two fermion bound state with the
corresponding dressing through scalar quanta and determine the same again
through energy extremisation. We also derive an effective two body potential
in this section in terms of a momentum dependent coupling
constant. In section {\bf 4} we illustrate the formalism with some
explicit calculations of the binding energy including renormalisation
effects. We also estimate the boson content of the bound state as a result
of dressing and the momentum fraction carried by these offshell bosons. We
tabulate the results to illustrate the effects of strong coupling. In
section {\bf 5} we discuss the results.

The present problem as a parallel of ``scalar" quantum chromodynamics
is meant to be an analogue for the consideration of hadrons.
We have consciously chosen a relatively simple problem
to illustrate the nonperturbative nature of the dynamics for
strong couplings. The results seem to reflect
features of gluon content of hadrons as observed in deep inelastic
collsions at the level of spectroscopy.
 \section{ \bf Fermion dressing}
We shall consider here the interacting fermions and hence as stated
the fermion will always be dressed with off-shell scalar quanta. We
shall consider here the same for a single fermion. We need this for
separating renormalisation effects.

The Hamiltonian we shall consider is given by
\begin{equation}
H=H_F +H_{INT} +H_{R}+ H_{CT},
\end{equation}
where,
$H_F$, the fermion kinetic term is given as
\begin{equation}
H_F=\int d{\vec z}\;c^{\dagger}(\vec z){\epsilon _z}c(\vec z).
\end{equation}
In what follows we shall take the fermions to be non-relativistic so that
$\epsilon _z=M-{{\vec \bigtriangledown}^{2} _{z}/ {2M}}$.
The interaction Hamiltonian $H_{INT}$ is given as
\begin{equation}
H_{INT}=g\int d\vec z\; c(\vec z)^{\dagger}
c(\vec z)\phi (\vec z).
\end{equation}
Further the free scalar field Hamiltonian $H_R$ is given as
\begin{equation}
H_{R}={1 \over 2}
\int d \vec z \; ({\dot \phi}^2 +(\vec \bigtriangledown \phi)^2+
\mu ^2 \phi ^2).
\end{equation}
 In the above $\mu $ is the scalar field mass.
Finally $H_{CT}$ will be the counter term to be identified with mass
renormalisation.

 We shall now expand the field
operators $\phi $ in terms of creation and annihilation
 operators as \cite{sah4,shut85,spm87}
\begin{eqnarray}
\phi(\vec z) & = & {1 \over {\sqrt{ 2 \omega _z}}}(a(\vec z )^{\dagger}+
a(\vec z)),\nonumber \\
\dot \phi (\vec z) & = &
i\sqrt {{\omega _z} \over 2}(-a(\vec z)+a(\vec z)^{\dagger}).
\end{eqnarray}

\noindent In the above $\omega _z$is a differentiation operator which e.g.
for free fields is given by\hfil
 ${(-{\vec \bigtriangledown_{z}}^2 +\mu ^2)}^{1/2}$. Further, here the meson
fields satisfy here the equal
time quantum algebra
\begin{equation}
\left[\phi (\vec x ),{\dot \phi} (\vec y)\right ]=
i\delta (\vec x -\vec y).\end{equation}
\noindent When we substitute the expansions for the field operators as in
equations (2) we have the usual commutation relations for the operators
$a$ and $a^{\dagger}$ as
\begin{equation}
\left[a(\vec x), a(\vec y)^{\dagger}\right]=\delta (\vec x -\vec y).
\end{equation}
\noindent Similarly for the fermion fields we have the anti commutation
relations given as

\begin{equation}
\left\{ c(\vec x),c(\vec y)^{\dagger}\right \}=
\delta (\vec x-\vec y).
\end{equation}

With the quantum algebra for the operators as above, we next wish to define
the ``dressed" particles as follows. We thus define
the ``physical" single fermion state as

\begin{equation}
\mid \vec x >={c^{phy}(\vec x)}^{\dagger} \mid vac> =
c(\vec x )^{\dagger} {U(\vec x)}\mid vac> ,\end{equation}
\noindent where the unitary operator $U(\vec x)$ is given as
\begin{equation}
{U(\vec x)} =exp(B(\vec x)^{\dagger}-B(\vec x)),
\end{equation}
with the operator $B^{\dagger}$ given as
\begin{equation}B^{\dagger}(\vec x)=\int d{\vec z}f(\vec x -\vec z){a(\vec
z)}^{\dagger}.
\end{equation}
\noindent Clearly $c(\vec x)^{\dagger}U(\vec x)\equiv
c^{phys}(\vec x)^{\dagger}$
is the creation operator with the fermions
being dressed. Thus the ``physical" fermion operator as we have defined
above, contains the bare fermion along with a coherent cloud of
scalar quanta. The distribution of these quanta is described through
the function $f(\vec x- \vec z)$ which shall be determined
through energy extremisation. We further note that from equations $(9)$
$(10)$ and $(11)$

\begin{eqnarray}
<\vec x\mid \vec y>& = & <vac\mid U(\vec x)^\dagger c(\vec x)
c(\vec y)^{\dagger}U(\vec y)\mid vac>\nonumber \\ & = &
\delta (\vec x -\vec y),
\end{eqnarray}
since $U(\vec x)$ is unitary and commutes with $c(\vec x)$. Hence
the physical single fermion state has the usual orthogonality
relation.

We shall now evaluate the hamiltonian expectation value
with respect to the single fermion state of equation $(9)$ which
after extremisation will determine the function $f$. Thus we
have
\begin{equation}<\vec x\mid H\mid \vec x'>= \delta (\vec x -\vec x')h(f),
\end{equation}
\noindent where
\begin{equation}h(f)=h_F +h_R +h_{INT}.\end{equation}

\noindent In the above $h_F$, $h_R$ and $h_{INT}$ correspond to
the expectation values of the hamiltonian $H_F$, $H_R$ and $H_{INT}$
of equations $(1)$ respectively and are given explicitly as
\begin{equation}
h_F= M+{1 \over {2M}}{1 \over (2 \pi)^3}\int d \vec k \; k^2
{{\tilde f}(\vec k)}^2,
\end{equation}
\begin{equation}
h_R={1 \over (2 \pi )^3}\int d \vec k\; {{\tilde f}(\vec k)}^2\omega (k),
\end{equation}
\noindent
and
\begin{equation}
h_{INT}={g \over (2 \pi)^3}\int d \vec k\; {\sqrt {2
\over \omega (k)}}{\tilde f}(\vec k).
\end{equation}
\noindent In the above $\tilde f$ is the Fourier transform of $f$
given through
\begin{equation}f(\vec x)={1 \over (2 \pi)^3}\int d \vec k {\tilde f}(\vec k)
e^ {i\vec k.\vec x}.\end{equation}
\noindent Extremising $h$ with respect to $\tilde f $ yields the
optimum $\tilde f$ as
\begin{equation}{\tilde f}(\vec k)=
-g \sqrt {2 \over \omega (k)}\left ({k^2 \over M}+
2 \omega (k)\right )^{-1}.\end{equation}
\noindent In the coordinate space the function $f(\vec x)$ is given as
using equation (11)
\begin{equation}
f(\vec x)=-g \sqrt{{2\over{\omega _x}}}\left(-
{{\vec \bigtriangledown_{x}}^2
\over M} + {2 \omega _x}\right )^{-1}.
\end{equation}

The field at a point $\vec z$ due to a coherent cloud of scalar
quanta associated with a fermion at $\vec x$ can be calculated as
\begin{equation}<\vec x\mid \phi (\vec z)\mid \vec x>=\sqrt{2 \over {\omega
_x}}f(\vec x
-\vec z)=-g {2\over \omega _x}(-{{\vec \bigtriangledown_{z}}^2 \over M}+
{2 \omega _x})^{-1}\delta (\vec x -\vec z).\end{equation}
\noindent Clearly in the limit of the fermion mass $M \rightarrow \infty$
equation $(21)$ reduces to the classical Yukawa field solution given as

\begin{eqnarray}
f(\vec x -\vec z)& = & -{g\over {\omega _x}^2}\delta (\vec x -\vec z)
\nonumber\\ & = & -\frac{g}{4\pi} e^{-\mu|\vec x-\vec z|}.\end{eqnarray}

\noindent Substituing the the expressions for ${{\tilde f}(\vec k)} $ in
equations (10) we have
\begin{equation}
h_F=M+{g^2 \over{2 \pi ^2}M }\int {k^4
 \over {\omega (k)(k^2 /M +2 \omega (k))^2}}\;dk,\label{eq:hf}
 \end{equation}
\begin{equation}
h_R={g^2 \over {\pi ^2}}\int {k^2
\over ({k^2 /M +2 \omega (k))^2}}\;dk ,\label{eq:hr}
\end{equation}

\noindent and
\begin{equation}
h_{INT}= -{g^2 \over {\pi ^2}}\int {k^2
\over {\omega (k)(k^2 /M +2 \omega (k))}}\;dk.
\label{eq:hi}\end{equation}
\noindent From the three contributions of equations $(23)$,$(24)$ and
$(25)$ above yields
the ``self energy" correction as
\begin{equation}\Delta M = h_F +h_{INT} + h_R-M
= -{g^2 \over {2 \pi ^2}}
\int {k^2 \over {\omega (k)(k^2 /M +2 \omega (k))}}dk.\end{equation}
\noindent Clearly the above is logarithmically divergent and hence
needs to be evaluated through a regularisation. We have to renormalise the
mass with a counter term in the Lagrangian/ Hamiltonian. Hence with
$\Delta M $ as in equation $(26)$, we shall now take
the counter term of equation (1) as
\begin{equation}
H_{CT}=-\Delta M\int c(\vec x)^{\dagger}
c(\vec x) d \vec x  .
\end{equation}

We note that we have for the single particle state in the above
included interactions through meson dressing as in equations (9) and (10) with
a specific form for the same as in equation (11).
$|\vec x>$ is not a single particle
eigenstate of the Hamiltonian, but an approximation of the same. Trial
states with all possible $B^\dagger$ shall generate the eigenstate.
$\tilde f(k)$ as determined in equation (19) gives an optimal approximation
for the form given by equation (11).

With the identification of the mass renormalisation as above with a coherent
dressing of the fermion we shall next consider the bound state of two
dressed fermions.

\section{\bf Two fermion bound state}

We shall define here the two fermion  bound state of total momentum
zero using the dressed fermions $c^{phys}$ of equation (9) as follows.
\begin{equation}\mid B(\vec 0)> =\frac{1}{(2 \pi )^{3/2}}\int u(\vec x -\vec
y)\;
 {c_1^{phys}}
(\vec x)^{\dagger}\;{c_2^{phys}}\;(\vec y)^{\dagger}\;
d {\vec x}d {\vec y}\mid vac>
,\end{equation}
\noindent Here
${c_i^{phys}}^{\dagger}(\vec x)$ (i= 1,2) has the same form as in equation
$(11)$.
Since we are considering the bound state, the distribution
of the scalar quanta around the fermions inside the bound state will be
different from that for single particle states. Thus we shall take
for the operator $B^\dagger (\vec x)$ of equation $(11)$ as
\begin{equation}B^{\dagger}(\vec x)=\int d \vec z \;
f_1 (\vec x -\vec z) a(\vec z)^{\dagger},
\end{equation}
\noindent where, $f_1$ is the dressing function of scalar quanta for
the ``constituent" fermions of the bound state and will be determined, as
before, through energy extreamisation. Further the function $u(\vec x -\vec y)$
is the conventional two particle wave funtion normalised as
\begin{equation}\int \mid u(\vec r)\mid ^2 d \vec r =1 .
\end{equation}
\noindent With two body wave function normalised as above, it is trivial
to check that $\mid B(\vec 0)>$ of equation $(28)$ is formally normalised as,
with equation (12),
\begin{equation}<B(\vec 0) \mid B(\vec 0)>=\delta (\vec 0).
 \end{equation}

Now we shall consider the expectation value of the Hamiltonian of equations
$(1)$ along with the counter term as given in equation (27)
with respect to the state as given in equation $(28)$. For the same we first
note that, using translational invariance, the energy expectation
value of the bound state will be given as \cite{spm87}
\begin{equation}h(f_1)=(2 \pi )^3 <B(\vec 0)\mid {\cal H}(\vec z)\mid B(\vec
0)>,
\end{equation}
\noindent where, ${\cal H}$ is the Hamiltonian density corresponding
to equations $(1)$ and $(27)$. Thus from the fermionic kinetic term
the contribution to the energy functional is given as
\begin{equation}h_F(\tilde f _1)=(2 \pi )^3 <B(\vec 0)\mid c_i
(\vec z)^\dagger(M-
{{\vec \bigtriangledown_{z}}^2 \over {2M}}) c_i(\vec z)\mid B(\vec 0)>
=2M +2 \Delta {M_F}_1 + T_F ,\end{equation}
\noindent where, $T_F$ is the conventional kinetic term given as
\begin{eqnarray}
T_F & = & \int u(\vec r)^{*} (-{{\vec \bigtriangledown_{r}}^2 \over M})
u(\vec r) d \vec r ,\nonumber\\
 & = & {1\over{(2\pi)^3}}\int {\tilde u}(\vec q)^* ({q^2\over {M}}){\tilde u}
 (\vec q)\; d\vec q.
 \end{eqnarray}
\noindent Further, $\Delta {M_F}_1$ is the contribution from
the scalar quanta dressing given as
\begin{equation}\Delta {M_F}_1 =\frac{1}{(2 \pi )^{3}} {1 \over 2M} \int k^2
{{\tilde f_1}(\vec k)}^2
d \vec k. \end{equation}
In the above ${{\tilde f_1}(\vec k)}  $ is the Fourior transform of
$f_1$ as in equation $(18)$. We note that this equation is
the second term in equation (15) where the fact that the dressing may change
when the fermion is a part of the bound state is included.

The contribution from the interaction term is given as
\begin{equation}
h_{int}(f_1)=(2 \pi )^{3 }<B(\vec 0)\mid g c^{\dagger} (\vec z)c(\vec z)
\phi (\vec z)\mid B(\vec 0)>
=T_{INT} + {\Delta M_{INT}}_1 ,
\end{equation}

\noindent where, parallel to equation $(17)$
\begin{equation}{\Delta M_{INT}}_1 = {1\over{(2 \pi )^3}} \int {\sqrt {2 \over
\omega (k)}}
{{\tilde f_1}(\vec k)} \; d \vec k .\end{equation}
\noindent $T_{INT}$ in equation $(36)$ is given as
\begin{equation}{T_{INT}(f_1)}= {4g \over (2 \pi )^3}
\int C(\vec q) {{{\tilde f_1}(\vec q)} \over {\sqrt{2 \omega (q)}}}\;
d \vec q ,
\end{equation}

\noindent where $C(\vec q)$ is related to the fermion wave function as
\begin{equation}C(\vec q)=(2 \pi )^{-3}\int {\tilde u}(\vec p +\vec q /2)
{\tilde u}(\vec p -\vec q /2) d \vec p .
\end{equation}

\noindent We have taken in the above ${\tilde u}(\vec k)$ as the Fourier
transform of $u(\vec r)$ defined as
\begin{equation}
u(\vec r)={1 \over {(2 \pi )^3}}\int {\tilde u}(\vec k)e^{i \vec k .\vec r}
\;d \vec k .
\end{equation}
\noindent Clearly, from the normalisation
of the two body wave function $C(\vec 0)=1$.  In a similar manner the
contribution from ${ H}_R$ is
\begin{equation}h_{R}= <B(\vec 0)\mid {\dot \phi }^2 - {\vec {\bigtriangledown
}
\phi }^2 + \mu ^2 \phi ^2 \mid B(\vec 0)> = T_R +2 {\Delta}M_{R_1}
,\end{equation}
\noindent where, $T_R$ is given as
\begin{equation}T_R= 2 \times (2 \pi)^{-3} \int {\tilde f_1}(\vec q)^2 C(\vec
q)
\omega (\vec q)\; d \vec q ,
\end{equation}
\noindent In the above $C(\vec q)$ is as defined in equation $(39)$ and
${\Delta M_R}_1$ is given as
\begin{equation}
{\Delta M_R}_1 =(2 \pi )^{-3} \int {{\tilde f_1}(\vec k)} ^2 \omega (k)
d \vec k .
\end{equation}

\noindent Thus the energy functional now becomes
\begin{equation}
h({\tilde f}_1)= T_F +T_{INT} +T_R +2M +2(\Delta M_1 -\Delta M),
\end{equation}
\noindent where,
\begin{equation}
\Delta M_1 ={\Delta M_F}_1 +{\Delta M_{INT}}_1 +{\Delta M_R}_1 ,
\end{equation}
and, $\Delta M $ term arises from the counter term as in equation $(27)$.
Extremising the energy
funtional $h({\tilde f}_1)$ with respect to the funtion ${\tilde f}_1$
yields the optimum
${\tilde f}_1$ as
\begin{equation}
{{\tilde f_1}(\vec k)}  = -g{\sqrt {2 \over {\omega (k)}}}
\times {(1+C(k))
\over {2C(k)\omega(k) +2 \omega (k) +{k^2} /M}}.
 \end{equation}
Substituting the above expression for ${{\tilde f_1}(\vec k)}$ in the
expression
for energy functional yields e.g.
$T_{INT}$ and $T_R$ as
 \begin{equation}
 T_{INT}=-{4 g^2 \over (2 \pi)^3}\int {C(\vec q) (1+C(\vec q))\over
 {\omega (q) \big [( q^2 /M +2 \omega (q) +2 C(\vec q)\omega (q)\big ]}}
 \;d\vec q,
 \end{equation}

 and,
 \begin{equation}
 T_R={4g^2 \over (2 \pi)^3} \int C(\vec q) \Bigl [{{1+C(\vec q)}\over
 {q^2 /M +2 \omega (q) +2 C(\vec q)\omega (q)}}\Bigr ] ^2 \;d
  \vec q .
  \end{equation}

 \noindent It can be seen that when ${{\tilde f}(\vec k)}$ is substituted in
 $\Delta M_1$ as in equation $(45)$ is also
 ultra violet divergent. However the quantity $\Delta M_1 - \Delta M$ does
 not have any ultra violet divergence and infact is given as
\begin{eqnarray} \Delta \epsilon & = & \big ( \Delta M_1 -\Delta M)\nonumber
\\&= &{g^2 \over (2 \pi)^3 }
 \int {{{({q^2 C(\vec q)/ M})}^2}\over
  {\omega (q)\big(q^2 /M +2 \omega (q)\big)
 \big(2C(\vec q)\omega (q) +q^2 /M + 2 \omega (q)}\big)^2}\;d\vec q.
\end{eqnarray}
 Thus the energy of the bound state is given as
 \begin{equation}
 E =2M +T_F + T_{INT} + T_R +2 \Delta \epsilon.
 \end{equation}
We note that through mass renormalisation we had in equation (49)
the {\em difference} of two divergent expressions which is convergent.
Here an additional contribution to binding energy arises due to
the  renormalisation effect.
This effect occurs in a nontrivial manner
through minimisation with respect to meson dressing.

A comment regarding renormalisation is needed. The present treatment
subtracts the one particle divergence in two particle interactions
through conventional counter terms. This is as per the philosophy of
renormalisation but is not equivalent to perturbative renormalisation.
 It is also incomplete since we do not know that it is applicable for
any general ansatz for the dressing of fermions.

 As may be noted in the above  the contributions $T_{INT}$, $T_R$
and $\Delta \epsilon $ constitute the potential energy. From the same
we shall identify the two body potential in the next subsection.

\subsection {\bf  Two body potential}
For the identification of the two body potential, we shall compare
the usual potential energy with energy expectation value as calculated here.
With a two body potential $V(\vec r)$ the corresponding
expression for the potential energy is given by

\begin{eqnarray}
V & = &\int u(\vec r)^* v(\vec r) u(\vec r)\;d \vec r ,\\
&  = &{1 \over (2 \pi)^3} \int C(\vec q) {\tilde v}(\vec q)\; d \vec q,
\end{eqnarray}
where $u(\vec r)$ is the two body wave function. In the second step
above we have written the same in the momentum space using equation $(40)$.
Further ${\tilde v}(\vec q)$ is the Fourier transform of $v(\vec r)$
defined through
\begin{equation}v(\vec r) = {1 \over (2 \pi )^3}
 \int {\tilde v}(\vec q)e^{i \vec k . \vec r}
\;d \vec q . \end{equation}

 From equation $(50)$ for the total energy, we also note that
subtracting the conventional kinetic energy $2M + T_F$ from the the total
energy will give the potential energy. We shall use this fact to define
the two body potential. Thus, the potential energy here arises from
$T_{INT}$, $T_R$ and $\Delta \epsilon$ of equations $(47)$, $(48)$
$(49)$ and, with equation $(52)$ the effective potential
 ${\tilde v}_{eff}(\vec q)$ gets defined through

\begin{equation}V = T_{INT} + T_R +\Delta \epsilon = {1 \over (2 \pi )^3}
\int C(\vec q) {\tilde v}_{eff} (\vec q)\; d \vec q ,
\end{equation}

\noindent where, the two body
effective potential ${\tilde v}_{eff}(\vec q)$ is given as
\begin{eqnarray}
{\tilde v}_{eff}(\vec q) & = &
-{g^2 \over {\omega (q)}^2 }\Biggl[{{4(1+C(\vec q))}
\over {(q^2 /M \omega (q) +2 +2 C(\vec q)}}-4 \bigg \{ {{(1+C(\vec q))}\over
{q^2 /M \omega (q) +2 + 2 C(\vec q)}}\bigg \}^2 \nonumber \\
& & -{C(\vec q){({q^2/M \omega (q)})^2}\over
{(q^2/M \omega (q) +2)\left(q^2 /M \omega (q) +2 + 2 C(\vec q)\right)^2}}
\Biggr].
\end{eqnarray}

\noindent We may rewrite the above equation as
\begin{equation}{\tilde v_{eff}(\vec q)}=-{{g_{eff}(\vec q)}^2 \over {\omega
(q)^2}},
\end{equation}
\noindent where  we have introduced a momentum dependent
effective coupling  ${g_{eff}}(\vec q)$ given as
\begin{mathletters}
\begin{equation}
{g_{eff}}^2 (\vec q) =  g^2 \times F(\vec q),
\end{equation}
where,
\begin{eqnarray}
F(\vec q) & = &
\Biggl[{{4(1+C(\vec q))}
\over {q^2 /M \omega (q) +2 +2 C(\vec q)}}-4 \bigg \{ {{(1+C(\vec q))}\over
{q^2 /M \omega (q) +2 + 2 C(\vec q)}}\bigg \}^2 \nonumber \\
& & -{C(\vec q) {({q^2/M \omega (q)})^2}\over
{(q^2/M \omega (q) +2)\left(q^2 /M \omega (q) +2 + 2 C(\vec q)\right)^2}}
\Biggr].
\end{eqnarray}
\end{mathletters}
The above equation may be seen as
the parallel of running coupling constant.
Clearly for soft processes i.e. $q \rightarrow 0$ we have $g_{eff}
\rightarrow g$ and ${\tilde V}_{eff} (\vec q)$ goes over to
the perturbative value $g^2 / \omega (q)^2$.
Also, when $q$ increases, $g_{eff}$ decreases to zero.
Such an idea of state and momentum dependent effective coupling constant is
purely of dynamical origin, and {\em does not} arise from any renormalisation
group equation.

Some comments regarding the definition of the potential as above
may be relevant. The potential as in equation $(55)$
contains fermion wave function ${\tilde u}
(\vec q)$ through $C(\vec q)$. The parameters of this function will get
determined through extremisation of energy of equation $(32)$. Then only
${\tilde V}_{eff}(\vec q)$ will be known after those parameters are substituted
in equation
$(55)$. The potential as defined above depends upon
the bound states of fermions and the fermion masses.
A contribution as above may be one of the reasons for the fermion mass
dependence of the potential in the consideration of
heavy quarkonium spectroscopy \cite{spm87,lichten}. In the
present picture of deriving the potential,
it happens to be an inevitable consequence of the formalism with
{\em simultaneous} minimisation over meson dressing and fermion wave function.

We could analytically do the extremisation of the energy functional here
because  the energy functional was quadratic in the function $\tilde f_1$
. This permitted an exact solution of the problem illustrating clearly
the physical conclusions. However, if there is a
quartic term $\lambda \phi ^4$ in the potential, or there is a cubic
and a quartic term, this extremisation $can not$ be explicitly done. However,
by choosing a suitable basis for the unknown functions, extremisation
can be done to a desired degree of accuracy. This opens up a new
frontier for the determination of effective potentials.

\section {\bf Some simple illustrations}

We shall here  first obtain the spin
indepedent central potential as varying with the bound state wave function.
For this purpose we take the
two body wave function in the harmonic oscillator wave function basis to
examine some features of the present nonperturbative approach. We take
two body fermion wave function as
\begin{equation}u(\vec r)= cos\!\alpha \, u_1(\vec r)
+ sin \!\alpha \,cos\! \beta \,u_2 (\vec r)
+sin\! \beta\,sin\!\alpha \, u_3(\vec r), \end{equation}
where,
\begin{eqnarray}
u_1(\vec r) & = & \bigg({1\over{\pi R^2}}\bigg)^{3/4}exp \bigg(-{r^2
\over{2R^2}}\bigg); \nonumber \\
u_2(\vec r) & = & \sqrt {3 \over 2}\bigg({1\over{\pi R^2}}\bigg)^{3/4}
\bigg(1-{{2r^2}\over {3R^2}}\bigg)
\exp \bigg(-{r^2 \over{2R^2}}\bigg);\nonumber\\
u_3(\vec r) & =
& {\sqrt{15 \over 8}}\cdot\bigg({1 \over {\pi R^2}}
\bigg)^{3 \over 4}
\bigg(1-{4r^2 \over {3R^2}}+4{r^4\over{15R^4}}\bigg)\exp \bigg(-{r^2
\over {2R^2}}\bigg).\end{eqnarray}
Clearly in the above we have taken the three terms of the basis and we shall
see  that it is sufficient to illustrate the results. With $u(\vec r)$ as above
$ C(\vec q)$
of equation $(39)$ becomes
\begin{equation}C(\vec q) = \sum {a_{ij}C_{ij}(\vec q)},
\end{equation}
where, $i,j=1,3$ and
\begin{equation}C_{ij}(\vec q)=(2 \pi )^{-3}\int {\tilde u}_i (\vec p +\vec q
/2)
{\tilde u}_j (\vec p -\vec q /2) d \vec p .
 \end{equation}

\noindent We may note that the two matrices $a$ and $C$ are symmetric
in their indices with the different elements given as
\begin{equation}
\begin{array}{lclclclcr}
a_{11} & = & cos^2\! \alpha , &  a_{12} & = & cos \alpha sin\!\alpha\,
cos\! \beta  , & a_{22} & =& sin^2\!\alpha \,cos^2 \!\beta
, \\
a_{13} & = & cos \! \alpha\, sin\! \beta\, sin \!\alpha ,&
a_{23}& = & sin^2 \alpha \, sin\!\beta\, cos \!\beta ,&
a_{33} & = & sin^2 \alpha sin^2 \beta
\end{array}
\end{equation}
and,
\begin{eqnarray}
C_{11}(\vec q)& = & exp(-{{q^2R^2}\over 4}),\nonumber \\
C_{12}(\vec q)& = & \sqrt{3\over 2}\cdot{1 \over 6}q^2 R^2
exp (-{q^2 R^2\over 4}),\nonumber \\
C_{22}(\vec q)& = & \big(1-{1 \over 3}
q^2 R^2 +{1\over 36} q^4 R^4 \big)exp (-{q^2 R^2 \over 4}),\nonumber \\
C_{13}(\vec q)& = & \sqrt{{15 \over 8}}\cdot{1 \over 60}\cdot q^4R^4 exp(-
{q^2R^2 \over 4}),\nonumber \\
C_{23}(\vec q)& = & {{3{\sqrt 5}\over 4}}\big({2 \over 9}q^2R^2 -{2\over
45}q^4R^4 +
{1 \over 360}q^6R^6\big)exp(-{q^2R^2\over4}),\nonumber\\
C_{33}(\vec q)& = &\big(1-{2 \over 3}q^2R^2 +{11 \over 60}q^4R^4 -{1 \over 60}
q^6R^6 + {1 \over 1920}q^8R^8\big)exp(-{q^2R^2 \over 4}).
\end{eqnarray}
With $C(\vec q)$ as in equation $(61)$, $T_{INT}$, $T_R$ and $\Delta \epsilon $
as given in equations $(47)$, $(48)$ and $(49)$ now become functions of
the fermion wave funtion parameters $\alpha $, $\beta$ and $R$.
 Further the fermion kinetic term $T_F$ now becomes
\begin{equation}T_F=\sum a_{ij}T_{ij},
\end{equation}
with,
$$T_{11}={3 \over 2}{1 \over {MR^2}};\quad
T_{12}=\sqrt{3 \over 2} {1 \over {MR^2}};\quad
T_{22}={7 \over 2} {1 \over {MR^2}};$$
$$ T_{13}=0 ;\quad T_{23}= \sqrt{5}\cdot {1 \over MR^2};\quad
T_{33}={11 \over 2}\cdot {1 \over MR^2}.$$

We next extremise the energy as given in equation $(50)$ with respsct
to the fermion wave funtion parameters $\alpha $, $\beta $ and $R$
for different
couplings and for different values of scalar field mass, which yields the
ground state energy.
 Once the parameters for the bound state are fixed
through extremisation, we may calculate the number of scalar
quanta in the bound state. This is given as

\begin{equation}N= <B(\vec 0)\mid{a^{\dagger}(\vec z)a(\vec z)}
\mid B(\vec 0)> =
2 \times (2 \pi)^{-3} \int {\tilde f_1}(\vec q)^2 C(\vec q)
\; d \vec q .
\end{equation}
As a post facto check for the nonrelativistic approximation for the
fermions one may calculate the average momentum of the fermions
{\em {inside the bound state}} which is given as
\begin{equation}
\frac{<{P_F}^2>}{M^2}  = {1\over{(2\pi)^3}}\int {\tilde u}(\vec q)^*
({q^2\over {M^2}}){\tilde u}
 (\vec q)\; d\vec q.
 \end{equation}
As may be noted from table I, this quantity increases as the coupling
is increased and illustrating how
the  nonrelativistic approximation for the fermions
in the bound state ceases to be valid.
We may include relativistic correction formally by replacing
the kinetic term contribution
 ${q^2/M}$ in equations $(34)$
and $(35)$ by $2((q^2 +M^2)^{1/2}-M)$ as well as in the solution for
${\tilde f}_1(k)$ in equation $(46)$. This replacement is not
exact, but it enables us to use the solution of equation (46)
 of the nonrelativistic limit to make a $qualitative$ extrapolation.

We next  calculate the momentum carried by the constituent
scalar quanta is given by
\begin{equation}
<P_S^2>=
<B(\vec 0)\mid {\vec {\bigtriangledown }a^{\dagger}}(\vec z)
 {\vec {\bigtriangledown }a(\vec z)}\mid B(\vec 0)>
=2 \times (2 \pi)^{-3} \int {\tilde f_1}(\vec q)^2
q^2 C(\vec q)
\; d \vec q .
\end{equation}
The results of the above calculations are summerised in Table I.

We note that the renormalisation corrections as calculated here
are consistently small. This is may be because we have constructed
the bound states with dressed fermions so that the main
effect of renormalisation has gone to define the physical mass.
 We also observe the nice result that the
deviations of the energy expectation values from the perturbative expressions
is not large. However, the distinction of the present method as opposed to
that of the potentials shows clearly in momentum distributions. We see that
as the coupling increases
from 0.1 to 1, the relative momentum fraction carried by the off-shell
bosons increases from about 10$\%$ to 67$\%$. This clearly illustrates
why gluons can carry about half the proton momentum in deep inelastic
lepton proton collisions. Such a result arises here through the
structure of the bound state.
We further note that in the last column
the average number of boson quanta increases from 6$\%$
when the coupling is 0.1 to as high as
288 for $g^2/4\pi=1$ and 3730 for $g^2/4\pi=1.5$
when we take $\mu=0$.
If we take the mass of the boson quanta to be 0.01M the average
number of these quanta rises from $3.5\times 10^{-5}$ for
coupling 0.1 to 1 and 3.4 for couplings 1 and 1.5 respectively.
The difference in these numbers only reflects the presence of soft
quanta for zero mass, and, illustrate that explicit gluon (or meson)
dressing should be relevant for properties of hadrons (or nucleus).

This picture for spectroscopy has specific extra predictions
for experimental observations. When we consider a probe which
interacts only with the fermions, the presence of off-shell
bosonic quanta simulating the potential will be known through momentum
imbalance, as seen in deep inelastic collisions. If the probe interacts
with boson quanta, they may also be directly ``seen" through these
interactions with the off-shell bosonic quanta of the
bound state. This could be operative for hadron hadron
collisions with the interaction of the gluons in hadrons.
In making the above statements we are obviously extrapolating
scalar mesons to that of gluons as a parallel
for hadrons in a realistic environment, or to pions in case of the
nucleus \cite{deut}.

Further, as the scalar field mass increases, the coupling
to form the bound state needs to be larger. For very strong
couplings ($g^2/4\pi> 2 $) we do not obtain a solution for the
bound state; the range of Yukawa interactions here becoming much too small.

We end our illustration with a possible example for Higgs particles.
We consider the case of heavy fermion  bound state
with Yukawa interaction where the coupling
becomes proportional to mass of the fermion as in grand unification
models \cite{inazawa88}.
Here we may have e.g. $g^2/4\pi=
{{\sqrt 2}G_F M^2 / 4\pi}$ with $G_F=1.7\times 10^{-5}GeV^{-2}$,
being Fermi coupling constant as in the minimal
Higgs model \cite{inazawa88}. The results here are
summarised in table II and the results are mostly similar to the results of
Table I. The difference in this case, however, is that
the momenta fraction carried by the off-shell
Higgs quanta is of the same order as that of fermions
even when the average number of
meson quanta is quite small.

\section{Discussions}

We would like to mention that the present paper is a part of
a new approximation scheme through a variational method in field
theory to vacuum structures or to bound states. For the later, the
scheme consists of defining dressed
single particle states \cite{deut} and then defining two particle bound
states through further modifications in dressing. Only equal
time algebra has been utilised. The energy of the one particle
or of the bound system is calculated through an extremisation procedure.
Since dressing is done through ansatz functions, a better
approximation will consist of taking more general functions.

The present paper shows that the effects of such ansatz functions
occur through highly nonlinear expressions which were not observed
in earlier formal analysis \cite{spm3}. As expected, the
one particle state contains the familiar divergence of self energy and
needs renormalisation. The use of renormalisation here follows
the familiar pattern of defining a counter term for a single particle
state so that the energy of the single particle at rest is identified with
its physical mass. We show that this counter term is adequate not to
have any infinity when the two particle bound state is defined. Also,
the counter term gives rise to a finite and state dependant additional
contribution to the energy of the bound state as expected
for any renormalisation procedure.

The prescription for renormalisation is still incomplete since
we do not show here that for {\em all} ansatz functions the
divergence disappears. We believe that
if infinities survive for physically measurable quantities, the solution
is likely to be inconsistent so that the ansatz function is not acceptable,
but the proof of the same shall be nontrivial.
Further we should probably include
a $\lambda \phi^4$ term as
a prototype of QCD to simulate the
quartic couplings of gluons. Then we may also have
bilinears in $\phi$ while considering dressing
\cite{ahss,ahss1,amspm,nm,nn,stev,reed,jphys}. However,
when these are included, the problem becomes prohibitively complicated.
In the present paper  therefore we have
implemented dressing to examine
renormalisation for single particle states so as to discuss
bound state spectroscopy related to other details of dynamics.

 This picture of including the scalar quanta in
  the bound state as off-shell constituents instead of using a propagator
  or potential is
not only aesthetically appealing but has phenomenological
consequnces for strong interactions as considered in the last section.
Also, the present formulation
generates state and fermion mass dependant potential as needed in quarkonium
spectroscopy, and, gives an insight in a
natural manner why gluons carry
about half the hadron momentum
in deep inelastic collisions. We can thus
ask and answer such questions while dealing with {\em spectroscopy}, and,
although really a toy model, as analysed
in the last section the results are
close to the expected physical situation.
\acknowledgements
The authors are thankful to A. Mishra, S.N. Nayak and P.K. Panda many
useful discussions. SPM would like to thank Department of Science and
Technology, Government of India for the research grant no.
SP/S2/K-45/89 for financial assistance.

\begin{table}
\squeezetable
\caption {}
\begin{tabular}{ccccccc}
\multicolumn{1}{c}{$g^2\over{4\pi}$} &\multicolumn{1}{c}{$E-2M$}
 &\multicolumn{1}{c}{$2\times \Delta \epsilon$}
 &\multicolumn{1}{c}{$E_{nonpert}/E_{pert}$}
  &\multicolumn{1}{c}{$<P_F^2>\over{M^2}$}
    &\multicolumn{1}{c}{$<P_S^2>\over{M^2}$}
 &\multicolumn{1}{c}{Avg. no. of scalar quanta} \\
\tableline
\multicolumn{7}{c}{Scalar field mass $\mu =0$}\\
\dec 0.01 &$-2.3\times 10^{-5}$
&$1.7\times 10^{-11}$  &1.00
&$2.4\times 10^{-5}$ &$2.3\times 10^{-7}$ &$1.5\times 10^{-4}$ \\
\dec 0.1 &${-2.3\times 10^{-3}}$
&$4.2\times 10^{-7}$
&${1.00}$ &$2.4\times 10^{-3}$
 &$2.1\times 10^{-4}$ &$6.1\times 10^{-2}$\\
\dec 1.0 &$-.244$ &$3.6\times 10^{-3}$ &$0.92$
 &$0.31$ &$0.20$ &$288$ \\
\dec 1.5 &$-0.618$ &$0.02$ &$0.81$ &$1.33$ &$1.02$
&$3730$\\
\\
\multicolumn{7}{c}{Scalar field mass $\mu=0.01M$}\\
\dec 0.1 &${-1.5\times 10^{-3}}$
&$5.2\times 10^{-7}$ &$1$
&$2.2\times 10^{-3}$
&$1.9\times 10^{-4}$ &$3.5\times10^{-5}$\\
\dec 1.0 &$-0.24$ &$3.8\times 10^{-3}$ &$0.92$
&$0.32$ &$0.2$ &1.01\\
 \dec 1.5 &$-0.61$ &$0.02$ &$0.81$ &$1.32$ &$1.02$ &$3.4$\\
\end{tabular}
\end{table}
\begin{table}
\squeezetable
\caption{}
\begin{tabular}{cccccccc}
\multicolumn{1}{c}{M (GeV)} &\multicolumn{1}{c}{coupling}
&\multicolumn{1}{c}{(E-2M)(Gev)}
&\multicolumn{1}{c}{${2\times \Delta \epsilon}$(GeV)}
 &\multicolumn{1}{c}{$E_{nonpert}/E_{pert}$}
   &\multicolumn{1}{c}{$<P_F^2>\over{M^2}$}
    &\multicolumn{1}{c}{$<P_S^2>\over{M^2}$}
&\multicolumn{1}{c}{Avg.No. of scalar quanta }
\\
\tableline
\multicolumn{8}{c}{$\mu =10 GeV$}\\
400 &0.21 &-2.33 & $3.5\times 10^{-3}$ &$0.99$ &$9.63\times 10^{-3}$
&$9.56\times 10^{-3}$ &$1.33\times 10^{-3}$\\
600 &0.47 &-26.9 &{0.13} &$0.97$ &$5.5\times 10^{-2}$ &$5.31\times 10^{-2}$
&0.132\\
800 &0.84 &{-128.2} &{1.5} &$.937$ &$0.206$ &$0.188$ &$0.607$\\
1000 &1.32 &{-440.32} &{12.0}
&$0.895$ &$0.777$ &$0.603$ &$2.21$\\
\\
\multicolumn{8}{c}{$\mu=100 GeV$}\\
600 &0.47 &{-2.38} &{0.03}
&$0.92$ &$2.85\times 10^{-2}$ &$2.80\times 10^{-2}$
&$1.3\times 10^{-2}$\\
800 &{0.84} &{-67.3} &{1.2} &$0.905$ &$0.183$ &$0.168$ &$0.184$\\
1000 &1.32 &{-333.1} &{11.3} &$0.828$ &$0.754$ &$0.588$ &$0.938$\\
\end{tabular}
\end{table}

\begin{references}
\bibitem{sah4} H. Mishra, S.P. Misra and A. Mishra,
Int.\ J.\ Mod.\ Phys. A3, 2331 (1988).
\bibitem{ahss} A. Mishra, H. Mishra, S.P. Misra and S.N. Nayak,
Pramana (Jou. of Phys) {37}, 59 (1991); {\em ibid} Z. Phys. C57, 233 (1993).
A. Mishra, H. Mishra and S. P. Misra, Z. Phys. C (To appear).
\bibitem{ahss1} A. Mishra, H. Mishra, S.P. Misra and S.N. Nayak,
Z. Phys. C57, 233 (1993);
H. Mishra and S. P. Misra  IP/BBSR/92-26.
\bibitem{amspm} A. Mishra and S.P. Misra, Z. Phys. C58,325 (1993).
\bibitem{nm} A. Mishra, H. Mishra,  and S.P. Misra
Int.\ J.\ Mod.\ Phys. A5, 3391 (1990);
H. Mishra, S.P. Misra, P.K. Panda and B.K. Parida,
Int. J. Mod. Phys. E1, 405 (1992).
\bibitem{nn} H. Mishra, S.P. Misra, P.K. Panda and B.K. Parida,
Int. J. Mod. Phys. E (To appear).
\bibitem{spm3}S.P. Misra, Phys.\ Rev.\ D{35}, 2607 (1987),
S.P. Misra, Ind.\ J.\ Phys. 61B, 287 (1987).
\bibitem{deut} S.P. Misra, P.K. Panda and R. Sahu, Phys. Rev. C45, 2079 (1992).
\bibitem{shut85} D. Schutte, Phys.\ Rev.\ D31, 810 (1985).
\bibitem{spm87} S.P. Misra, S. Naik and A.R. Panda,
 Pramana (J. Phys.) 28, 131 (1987).
\bibitem{lichten}D.B. Lichtenberg, E. Predenzzi, R. Roncaglia, M. Rosso and
J.G. Wills, Z. Phys. C41, 615 (1989).
\bibitem{inazawa88} H. Inazawa and T. Morri, Phys.\ Lett.\ {B203},
 279 (1988).
\bibitem{stev}P. M. Stevenson, Phys.\ Rev.\ D {32}, 1389 (1985);
\bibitem{reed}P. M. Stevenson, G.A. Hajj and J.F. Reed, Phys. Rev. D34,
\bibitem{jphys}H. Mishra and A. R. Panda, J. Phys. G (Nucl. Part. Phys.) 18,
1301 (1992).
\end{references}
\end{document}